\documentclass{aa}  

\usepackage{amsmath} 
\usepackage{graphicx}
\usepackage{txfonts}
\usepackage{soul}
\usepackage{xcolor}

\usepackage{subcaption}
\begin{document}

   \title{AutoSourceID-Light}

   \subtitle{Fast optical source localization via U-Net and Laplacian of Gaussian}

   \author{F. Stoppa \inst{1,7}
        \and
        P. Vreeswijk \inst{1}
        \and
        S. Bloemen \inst{1}
        \and
        S. Bhattacharyya \inst{2}
        \and
        S. Caron \inst{3,4}
        \and
        G. Jóhannesson \inst{5}
        \and
        R. Ruiz de Austri \inst{6}
        \and
        C. van den Oetelaar \inst{3}
        \and 
        G. Zaharijas \inst{2,13}
        \and 
        P.J. Groot \inst{1,8,9,10}
        \and
        E. Cator\inst{7}
        \and
        G. Nelemans \inst{1,11,12}
      }

   \institute{
            Department of Astrophysics/IMAPP, Radboud University, PO Box 9010,
            6500 GL, The Netherlands
            \and
            Center for Astrophysics and Cosmology, University of Nova Gorica, Vipavska 13, SI-5000 Nova Gorica, Slovenia
            \and
            High Energy Physics/IMAPP, Radboud University, PO Box 9010,
            6500 GL, The Netherlands
            \and
            Nikhef, Science Park 105, 1098 XG Amsterdam, the Netherlands
            \and
            Science Institute, University of Iceland, IS-107 Reykjavik, Iceland
            \and
            Instituto de Física Corpuscular, IFIC-UV/CSIC, Valencia, Spain
            \and
            Department of Mathematics/IMAPP, Radboud University, PO Box 9010, 6500 GL Nijmegen, The Netherlands
            \and
            Department of Astronomy, University of Cape Town, Private Bag X3, Rondebosch, 7701, South Africa
            \and
            South African Astronomical Observatory, P.O. Box 9, Observatory, 7935, South Africa
            \and
            The Inter-University Institute for Data Intensive Astronomy, University of Cape Town, Private Bag X3, Rondebosch, 7701, South Africa
            \and
            SRON, Netherlands Institute for Space Research, Sorbonnelaan 2, NL-3584 CA Utrecht, The Netherlands
            \and
            Institute of Astronomy, KU Leuven, Celestijnenlaan 200D, B-3001 Leuven, Belgium
            \and 
            Institute for Fundamental Physics of the Universe, Via Beirut 2, 34151 Trieste, Italy
            }

\date{\today}

  \abstract
   {}
   {With the ever-increasing survey speed of optical wide-field telescopes and the importance of discovering transients when they are still young, rapid and reliable source localization is paramount. 
   We present AutoSourceID-Light (ASID-L), an innovative framework that uses computer vision techniques that can naturally deal with large amounts of data and rapidly localize sources in optical images.
  }
   {We show that the ASID-L algorithm based on U-shaped networks and enhanced with a Laplacian of Gaussian filter provides outstanding performance in the localization of sources.
   A U-Net network discerns the sources in the images from many different artifacts and passes the result to a Laplacian of Gaussian filter that then estimates the exact location. 
   }
   {Using ASID-L on the optical images of the MeerLICHT telescope demonstrates the great speed and localization power of the method. We compare the results with SExtractor and show that our method outperforms this more widely used method. ASID-L rapidly detects more sources not only in low- and mid-density fields, but particularly in areas with more than 150 sources per square arcminute. The training set and code used in this paper are publicly available.}
   {}


   \maketitle
%

\setlength{\parindent}{0em}

\section{Introduction}


The increased capabilities of many telescopes both on Earth, for example the Zwicky Transient Facility (ZTF, \citealp{bellm2014, bellm2019}) and the Vera C. Rubin Observatory \citep{Ivezi2019}, and in orbit, such as Gaia \citep{Gaia2016} and the recently launched James Webb Space Telescope (JWST, \citealp{Gardner2006}), are and will be able to provide large amounts of data at a staggeringly increasing rate. Synoptic telescopes can already take images of the size of $100$ Mpixels every $15-60$ seconds; this processing speed creates a real bottleneck, and thus there is a need for analysis techniques that can efficiently keep up with this trend and can naturally deal with large amounts of data.

\parskip = 0pt
There is a long list of methods used in the astronomical community for source localization. Among the most famous are SExtractor \citep{Bertin1996}, Daophot \citep{Stetson1987}, Mopex \citep{Makovoz2005}, SourceMiner \citep{Savage2007}, and Astrometry.net \citep{Lang2010}. Most of these methods use a combination of image transformation and detection criteria to first estimate the background and then perform thresholding and deblending to separate overlapping sources. However, when the density of sources exceeds a certain level, these multi-step processes can struggle and their performance can degenerate both in computation time and localization accuracy.

In this paper we propose a new way to analyze optical imaging data that uses computer vision techniques to rapidly localize sources. This method is also extendable to different parts of the electromagnetic spectrum, and therefore naturally leads to the possibility of simultaneous multi-wavelength source analysis. \\
Some of us developed the AutoSourceID (ASID) code, targeted at automatic sources localization and classification in gamma-ray data \citep{Panes2021}.
ASID showed significant promise; the source detection threshold was comparable to that of the traditional catalogs (e.g., 4FGL) \citep{Abdollahi2020}, but with the added advantage that source detection proved to be more robust to uncertainties of the diffuse gamma-ray background. 

The code presented in this paper uses a similar baseline structure and is focused on the rapid localization of sources in optical images, the reason behind naming it AutoSourceID-Light (ASID-L). 
ASID-L uses a U-Net \citep{Ronneberger2015} network to construct a segmented mask where each pixel in the input image is assigned a value between 0 (background) and 1 (source). A Laplacian of Gaussian filter \citep{Chen1987} is then applied to the mask predicted by the U-Net to identify the individual sources. The result is a fast automatic way to go from images to a catalog of sources.

Moreover, the problems encountered by the previously mentioned source detection methods can be solved by deep learning algorithms whose computation time does not depend on the number of sources in the images and which, if well trained, can retrieve more sources in crowded regions.

In this work we use wide-field optical images taken with the MeerLICHT telescope, a 65 cm telescope located in Sutherland, South Africa \citep{BloemenGroot2016,Groot2019}. \\
To train the U-Net we also need the true locations of the sources in the images; since the main interest in the context of MeerLICHT is the rapid localization of point sources, we retrieved these locations from the Gaia Early Data Release 3 catalog (EDR3, \citealp{Gaia2016,Gaia2021}). As in the case of any supervised machine learning algorithm, the choice of the training set is fundamental and deeply related to what the network will learn; in our case this means that only point sources will be recognized by the network as targets for its localization. The ASID-L framework is not limited to point sources, however; the U-Net creates a circular segmentation mask for each source that it is trained to localize regardless of its shape, opening up the possibility of localizing extended sources such as galaxies if they are part of the training process.

Finally, we compare the results of ASID-L with SExtractor \citep{Bertin1996}. We chose this method not only because it is widely used in the community, but also because it is already part of the optical images processing pipeline of the MeerLICHT telescope.

\section{Method}

ASID-L, like its gamma-ray counterpart ASID, works with a chain of networks and tools to rapidly create a catalog of sources from an image. 
For optical images, the pivotal steps are to detect and find the precise location of the sources; therefore, ASID-L is divided into the following steps: mask generation and source localization. 
We describe each step in more detail below.

\subsection{Mask generation: U-Net image segmentation} 

Evolved from the well-known convolutional neural network (CNN) architecture (\citealp{LeCun1999}), U-Net was first designed and applied in 2015 to process biomedical images. 
The U-Net architecture is primarily used for semantic segmentation, where for each pixel of an image, a corresponding class label is predicted, and helps provide the pixel-level contextual information. Apart from biomedical image segmentation, U-Net was also applied to a wide range of case studies such as lunar crater detection \citep{Wang2020}, radio astronomy \citep{akeret2017}, and cosmology \citep{Bonjean2020}.

U-Net follows a symmetric encoder--decoder structure. In the encoding path, the image dimension is halved at each stage, and the number of channels are increased in the convolution operation. The channels can be thought of as feature detectors where, low-level features are detected in the initial part of the network and high-level features are detected as the image size reduces and thus we can obtain a dense feature map. This part of the network is similar to a regular fully CNN (FCN \citealp{long2015}) and provides classification information.

The obtained dense feature map is then increased back in size in the decoder path to obtain the full-scale segmented image, which is the reason behind the ``U'' in the network name. In the decoder path, each convolutional layer is connected to its equal-sized counterpart of the encoding path, which helps to combine the small-scale and the large-scale structure and propagate the contextual information along the network to make accurate small-scale predictions on the obtained final segmentation map. \\
During the training of the network, the weights and biases of the convolutional kernels are optimized based on the loss function described in Section \ref{Loss function}.

The input for the training set of a U-Net algorithm are the images to be segmented and their known segmentations, called masks. After the training process, the U-Net only needs an image as input and will predict its segmentation mask.   
U-Net networks are typically used to cut out a few relatively large structures (e.g., in biology or galaxy images); our use-case, discerning many small objects, is largely unexplored in fundamental science.

In our application to MeerLICHT, the input images for the training set are pairs of $256 \times 256$ pixel patches obtained from full field images and their corresponding masks. More details about the location of the sources and the choice of a fixed size mask is in Section \ref{Training set}.

\begin{figure}[h!]
     \centering
     \begin{subfigure}[b]{0.24\textwidth}
         \centering
         \includegraphics[width=\textwidth]{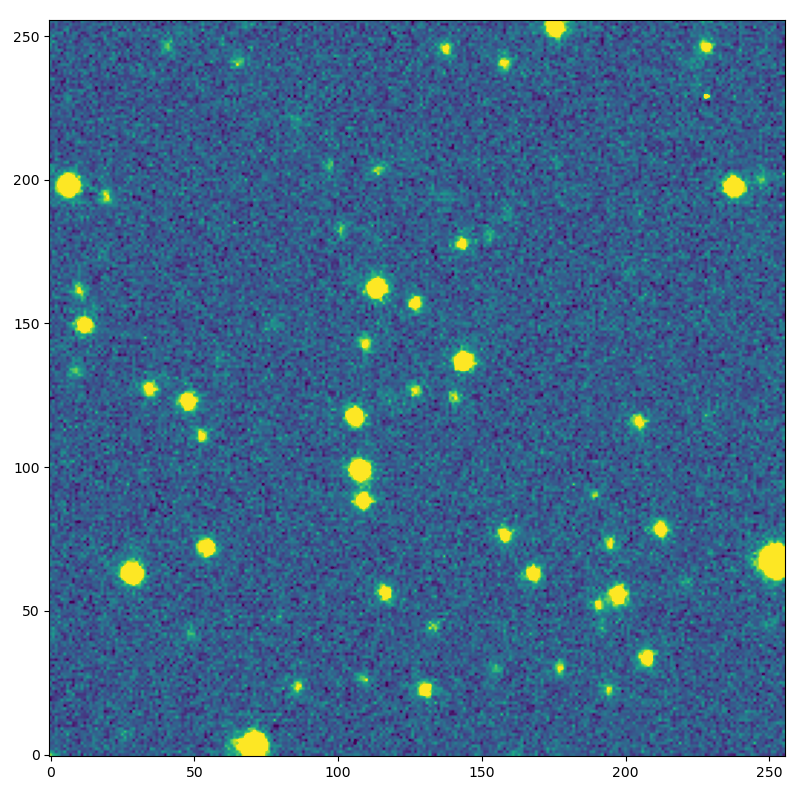}
         
     \end{subfigure}
     \hfill
     \begin{subfigure}[b]{0.24\textwidth}
         \centering
         \includegraphics[width=\textwidth]{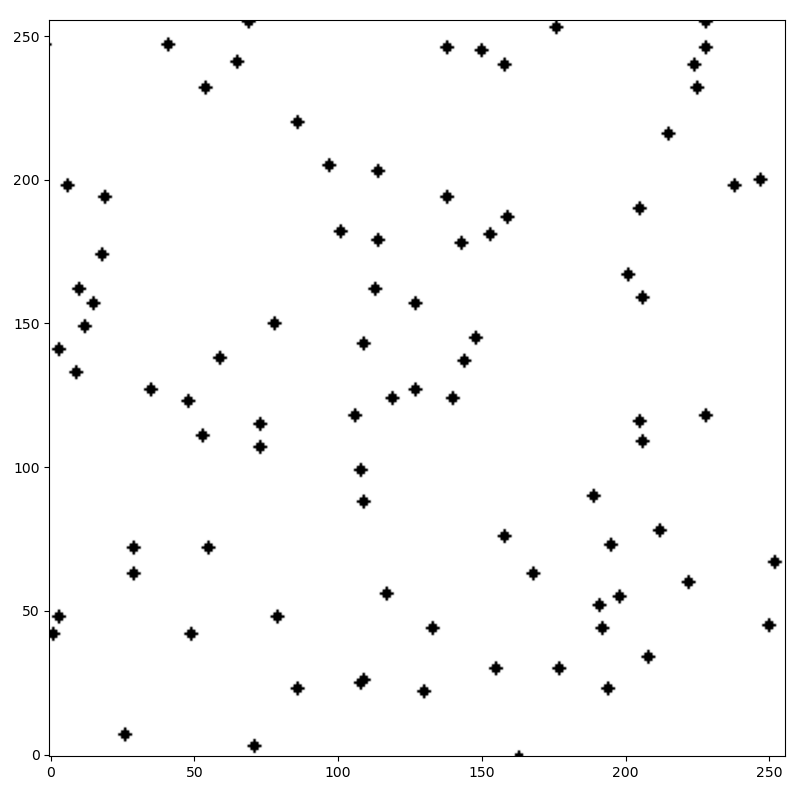}
    \end{subfigure}
     
\caption{U-Net input, optical image patch on the left and associated mask patch on the right.}
\label{fig: input patch}
\end{figure}

\begin{figure}[h!]
\centering
\includegraphics[width=0.4\textwidth]{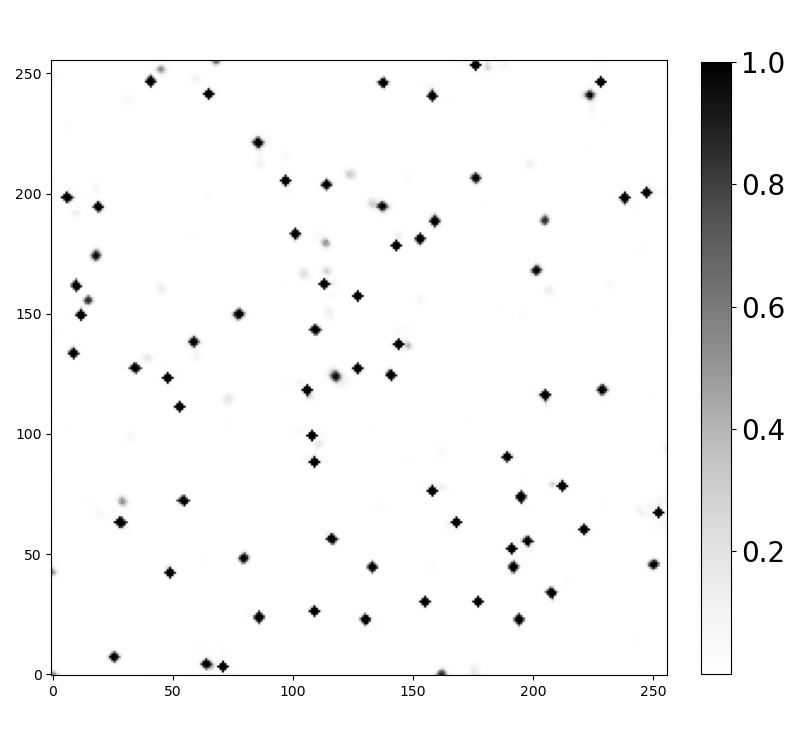}
\caption{U-Net output, predicted mask for the left image of Fig. \ref{fig: input patch}.}
\label{fig: output patch}
\end{figure}

The training process is straightforward. At epoch 1 the U-Net is fed with thousands of pairs of optical images and their associated masks; an example is shown in Fig. \ref{fig: input patch}. Starting only from the optical image of the pair, the U-Net predicts a mask similar to the one shown in Fig. \ref{fig: output patch}. In reality, at epoch 1 the U-Net does not know anything about what a mask looks like, so the result would be much worse than that shown in Fig. \ref{fig: output patch}. Comparing predicted and training masks with the loss function of Section \ref{Loss function}, the U-Net gradually learns where and how to improve its prediction for the next epoch. By repeating this process for multiple epochs, the U-Net is able to reconstruct mask patches that closely resemble those in the training set, such as the one shown in Fig. \ref{fig: output patch}.

\noindent
The main difference between a training mask and the output of the U-net is that in the latter the predictions are continuous values in the range $[0,1]$ for each pixel (instead of integers $\{0,1\}$). Thus, the need for an additional method to discern which groups of pixels should be considered sources and where their centers are located.

\subsection{Loss function}
\label{Loss function}

To optimize the model during training, we applied a combination of binary cross-entropy (BCE, \citealp{Mannor2005}) loss and Dice \citep{Sudre2017} loss.

Binary cross-entropy loss is defined as
\begin{equation}
\text{BCE loss} = - \frac{1}{n}  \sum_{i=1}^{n} y_{\text{true}, \, i} \, log(y_{\text{pred}, \, i}) + (1-y_{\text{true}, \, i}) \, log(1-y_{\text{pred}, \, i}) 
\label{eqn:BCE}
.\end{equation}

\noindent
This loss examines each pixel individually, comparing $y_{\text{true}}$ (the true mask value, either 0 or 1) assigned to a specific pixel to $y_{\text{pred}}$ (the mask value predicted by the U-Net), and then averages over all $n=256\times256$ pixels.

The second loss function is based on the Dice coefficient \citep{Dice1945}, which is a measure of the overlap between two samples. 
In 2017 it was adapted as a loss function known as Dice loss, and it is defined as

\begin{equation}
\text{Dice  loss} = \frac{2 \sum_{i=1}^n y_{\text{true}, \, i} \;  y_{\text{pred}, \, i}} {\sum_{i=1}^n y_{\text{true}, \, i}^2  + \sum_{i=1}^n y_{\text{pred}, \, i}^2  }
\label{eqn:Dice loss}
.\end{equation}

The BCE loss works best for equal data distribution among classes, while the Dice loss is particularly suitable for segmentation tasks. The combination of these two losses, also known as Combo loss \citep{taghanaki2021}, was shown to improve performance in medical image segmentation tasks with a class-imbalanced dataset. Inspired by this, we employed in ASID-L the sum of these two losses as the loss function.

\subsection{Source localization: Laplacian of Gaussian}
\label{Source localization}

Once we have a predicted mask like that in Fig. \ref{fig: output patch}, we apply the Laplacian of Gaussian (LoG) algorithm to determine the exact location and number of sources.

The LoG is a computer vision method for the detection of blobs, based on the combination of a Laplacian filter and a Gaussian blurring step \citep{Sotak1989, Lindeberg1992}.
A Laplacian filter is a derivative filter used to find areas of rapid change in images and, for an image with pixel intensity values $I(x,y)$, is given by

\begin{equation}
\nabla^2 I(x,y)=  \frac{\delta^2 I(x,y)}{\delta x^2} \frac{\delta^2 I(x,y)}{\delta y^2}.
\label{eqn: laplacian}
\end{equation}

\noindent
Since the Laplacian is a second-order derivative filter, it is quite sensitive to noise. For this reason a Gaussian blurring step is needed to mitigate the problem.
Both the Gaussian kernel and the Laplacian filter can be applied simultaneously due to the associative property of the convolution operation, and this two-in-one process is called LoG. The result is a great tool for blob localization that identifies regions that differ in properties from surrounding areas.

A 2D LoG function centered on zero and with standard deviation $\sigma$ has the form 
\begin{equation}
    \text{LoG}(x,y; \sigma^2) =-\frac{1}{\pi \sigma^4} \left[ 1 - \frac {x^2 + y^2 }{2 \sigma^2} \right]  e^{- {\frac {x^{2}+y^{2}}{2\sigma^2} }}. 
\label{eqn: Laplacian of Gaussian}
\end{equation}

\begin{figure}[h]
\centering
\includegraphics[width=.4\textwidth]{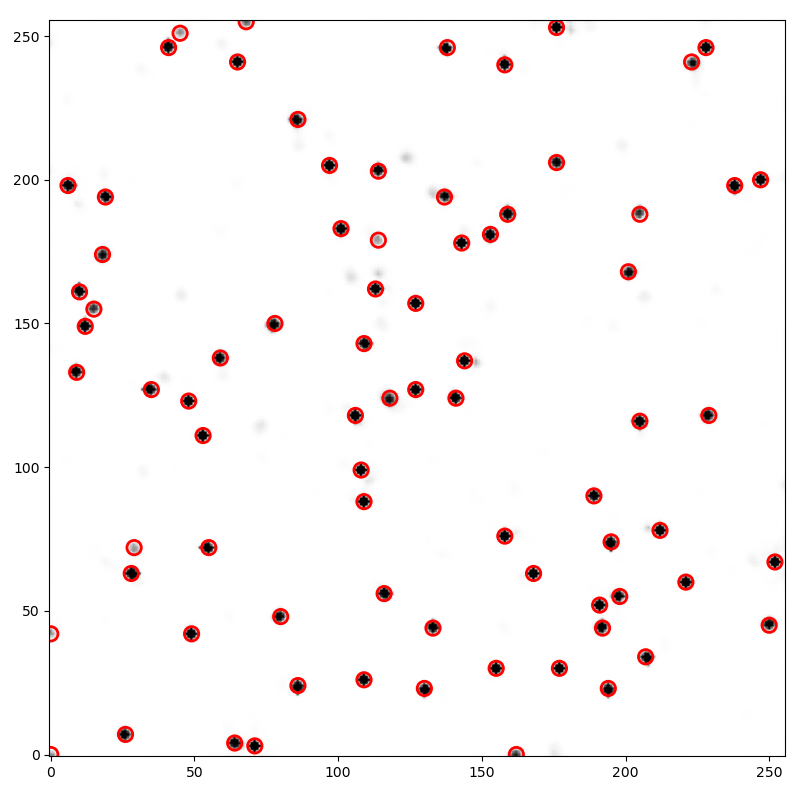}
\caption{Sources localized by the LoG (red circles) in the U-Net predicted mask.}
\label{fig: LoG}
\end{figure}

\noindent   
The LoG is strongly dependent on the choice of $\sigma$ due to the relationship between the size of the blob structures in the image and the Gaussian kernel. As constructed, the LoG results in high positive values for blobs of radius close to ${\sqrt {2}\sigma}$. In general, to capture blobs of different sizes a multi-scale approach is needed where the LoG is applied multiple times with a set of different values for $\sigma$ \citep{Lindeberg1998, Lindeberg2013}. However, this is not the case for us; in fact, we exploit the fixed size of our masks to find the single best $\sigma$ to localize our blobs, improving both accuracy and speed of ASID-L (for more details, see Section \ref{LoG parameters}). Figure \ref{fig: LoG} shows an example of what the LoG filter identifies as sources and their locations for the predicted mask of Fig. \ref{fig: output patch}. 

\begin{figure}[h]
\centering
\includegraphics[width=.4\textwidth]{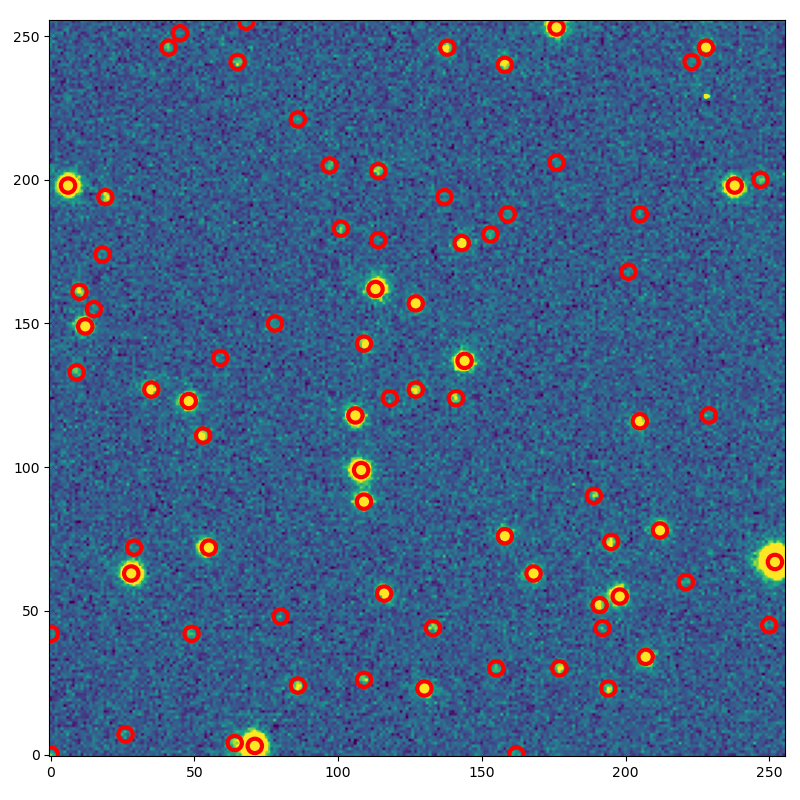}
\caption{Sources localized by ASID-L (red circles) superimposed on the optical image.}
\label{fig: LoG on Optical}
\end{figure}

\noindent
This is the final output of ASID-L, the locations identified by the LoG are reported as a list of pixel coordinates, but can easily be converted to any coordinate system. We can see the results of ASID-L superimposed on the original optical image in Fig. \ref{fig: LoG on Optical}.

\section{Application}

We now show how we built ASID-L on images taken with an optical telescope, in our case the MeerLICHT telescope \citep{BloemenGroot2016,Groot2019}. 

\subsection{MeerLICHT}
\label{MeerLICHT}

The MeerLICHT telescope is a 65 cm optical telescope with a field of view of 2.7 square degrees and a 10.5k $\times$ 10.5k pixel CCD.
\noindent
The main aim of MeerLICHT is to follow the pointings of the MeerKAT radio telescope \citep{Jonas2016} to enable the simultaneous detection of transients at radio and optical wavelengths. The filter set available is the SDSS $ugriz$ set and an additional wide $g$+$r$ filter named $q$. The images taken are immediately transferred to the IDIA/ilifu facility, where the image processing software BlackBOX (https://github.com/pmvreeswijk/BlackBOX, Vreeswijk et al., in prep) processes the images in the standard fashion before continuing with the source detection (currently using SExtractor), the astrometric and photometric calibration, the derivation of the position-dependent image point spread function (PSF), the image subtraction, and transient detection. \\
The code presented here is the second deep learning algorithm developed in the context of MeerLICHT, following MeerCRAB, an algorithm used to classify real and bogus transients in optical images \citep{Hosenie2021}.

\subsection{Training set}
\label{Training set}

To build and evaluate ASID-L, we selected MeerLICHT $q$-band images of fields with different source densities: (1) a field centered on the Omega Cen globular cluster, (2) a field of the Fornax galaxy cluster and (3) an ``empty'' field centered on the Chandra Deep Field-South (CDF-S, \citealp{Giacconi2002}). We used the Gaia Early Data Release 3 catalog (EDR3, \citealp{Gaia2016,Gaia2021}) to infer the presence of real sources for the training set. For each of the above fields we selected the relevant Gaia EDR3 sources and converted the G-band magnitude of the Gaia source to the q-band flux (in electrons per second) that the source would have on a specific image, using the image zero-point determined in the MeerLICHT photometric calibration. Together with the image background noise, consisting of a combination of the sky background noise and the read noise, and the PSF shape at the source position on the image, we were then able to determine the signal-to-noise ratio (S/N) that the Gaia source would have on a specific MeerLICHT image.
The number of sources per square arcminute as a function of the S/N for each field is shown in Table \ref{table:1}.

\begin{table}[h]
\caption{Number of sources with different S/N thresholds}             
\label{table:1}      
\centering                          
\begin{tabular}{c c  c c c}        
\hline\hline                 
Field & $S/N\geq1$ & $S/N\geq3$  & $S/N\geq5$ & $S/N\geq7$ \\    
\hline                        
   (1) & 38.304 & 37.002 & 34.856 & 32.740 \\
   (2) & 0.889 & 0.816 & 0.710 & 0.621 \\
   (3) & 0.929 & 0.810 & 0.684 & 0.590 \\
\hline                                   
\end{tabular}
\tablefoot{
Number of sources per square arcminute in the three fields: (1) Omega Centauri cluster, (2) Fornax galaxy cluster, and (3) ‘`Empty'’ CDF-S
}
\end{table}

\noindent
Variability in the brightness of the point sources and colour terms (MeerLICHT $q$-band is much narrower than Gaia $G$) lead to differences between predicted S/N of a source based on Gaia and the true S/N in the MeerLICHT image. The sharp cut-off at the predicted S/N in reality becomes a soft cut-off around that S/N. 

For the training, test, and validation sets, we used the three fields described above and the masks built from Gaia sources with a S/N above 3. In Section \ref{SNR} we come back to our S/N cut-off choice. 

Each field is divided into 1681 patches of $256 \times 256$ pixels, for a total of 5043 patches of optical images. The choice behind the size of the patches was determined by having a reasonable number of trainable parameters in the U-Net, approximately two million, and at the same time enough memory to load the images. With dedicated hardware the size of the images can be increased.
We then created a field mask for each field; at every Gaia EDR3 source location we created a mask made of a central $3\times3$ square of pixels and an additional pixel in every cardinal direction. This is the smallest number of pixels such that the mask still resembles a circle and for overlapping to be minimized in very crowded regions. 
Each field mask was then split in the same way as for the optical images, resulting in 5043 $256 \times 256$ mask patches. 
We assigned $80\%$ of the patches to the training set, $10\%$ to the test set, and $10\%$ to the validation set.

\section{Results}

To evaluate the performance of ASID-L in terms of localized and non-localized sources with respect to the Gaia EDR3 catalog we use the Dice coefficient.
In confusion matrix settings, the Dice coefficient can be framed via the following formula:

\begin{equation}
\text{Dice}  ={\frac {2 TP}{2 TP+FP+FN}}
\label{eqn:Dice coeff2}
.\end{equation}

\noindent
Here true positive (TP) is the total number of sources that are both localized by ASID-L and in the Gaia catalog, false positive (FP) is the total number of sources where ASID-L localized something that was not in the Gaia catalog, and false negative (FN) is the total number of sources missed by our method that instead were in the Gaia catalog.

In particular, we evaluate the results of ASID-L on the $165$ test patches belonging to the Omega Cen globular cluster; we chose this specific field due to the high variability in the number of sources in each patch. Outside the cluster, an average of a few dozen sources can be identified in each patch, while closer to its center hundreds or even thousands of sources can be found in a $256 \times 256$ patch.

\subsection{S/N cut-off choice}
\label{SNR}

The choice of the S/N cut-off for the training set locations influences the results of ASID-L and has to be well thought out. A cut-off that is too high in Gaia EDR3 means losing sources that are actually in the images, and consequentially, ASID-L will learn to ignore anything below the cut-off. On the contrary, a cut-off that is too low will create masks in locations with no sources in the optical image resulting in a very high number of FPs in the output of ASID-L. In Table \ref{table:2} we report the number of localized sources per square arcminute as a function of the S/N cut-off of the training set; based on this information we chose the S/N cut-off that results in the highest Dice coefficient and lowest S/N: $S/N=3$.

\begin{table}[h]
\caption{Training results with different S/N cut-off}             
\label{table:2}      
\centering                          
\begin{tabular}{c c c c c c}        
\hline\hline                 
S/N & TP & FN & FP & Dice   \\    
\hline                        
  1     & 26.887 & 10.071 & 2.823 & 0.8066 \\
  \textbf{3} & \textbf{27.001} & \textbf{9.957} & \textbf{2.918} & \textbf{0.8075}  \\
  5     & 25.898 & 11.060 & 2.031 & 0.7983  \\
  7     & 24.620 & 12.338 & 1.743 & 0.7776  \\
\hline                                   
\end{tabular}
\tablefoot{Number of TPs, FNs, and FPs per square arcminute detected as a function of the S/N cut-off of the training set for the 165 patches of the Omega Cen test set. Boldface indicates the chosen S/N cut-off.}
\end{table}

\subsection{LoG parameters choice}
\label{LoG parameters}

There are three parameters to optimize for our LoG step: the standard deviation $\sigma$, a threshold $\tau$, and an overlap index $\omega$. 

As introduced in Section \ref{Source localization}, we exploited the fixed size of the masks to estimate the optimal $\sigma$ parameter. Evaluating multiple values of $\sigma$ we found that $\sigma=1.43$ is the optimal choice to localize blobs of the exact size and shape of our masks. As a consequence, we improved the accuracy of the LoG ensuring that only the U-Net's predicted blobs of the correct size are localized and substantially increasing the speed by not having to evaluate multiple values of $\sigma$ at every iteration. 

To simplify the job of the LoG, a threshold $\tau$ is set to remove any predicted pixel with a value below it. An immediate way to see the influence of the threshold $\tau$ on the LoG result is the number of FPs. The information in Table \ref{table:3} suggested that a LoG threshold of $0.2$ is the optimal choice.\\

\begin{table}[h]
\caption{Detected sources with different LoG thresholds.}             
\label{table:3}      
\centering                          
\begin{tabular}{c c c c c }        
\hline\hline                 
Threshold & TP & FN & FP & Dice \\    
\hline                        
0.00 & 30.371 & 6.586 & 109.651 & 0.3432 \\
0.05 & 28.501 & 8.456 & 7.805 & 0.7780 \\ 
0.10 & 27.922 & 9.035 & 4.946 & 0.7998 \\ 
0.15 & 27.452 & 9.506 & 3.666 & 0.8065 \\
\textbf{0.20} & \textbf{27.001} & \textbf{9.957} & \textbf{2.918} & \textbf{0.8075} \\
0.25 & 26.542 & 10.416 & 2.351 & 0.8061 \\
0.30 & 25.867 & 11.091 & 1.921 & 0.7990 \\

\hline                                   
\end{tabular}
\tablefoot{Number of TPs, FNs, and FPs per square arcminute detected as a function of the LoG threshold for the 165 patches of the Omega Cen test set. Boldface indicates the chosen threshold cut-off.}
\end{table}

\noindent
The last parameter is the maximum amount of overlap $\omega$ between adjacent sources. We set this parameter to $0.8$, meaning that if two adjacent sources overlap by more than $80\%$ only one central source is localized. For our test set, this parameter has no influence on the resulting number of TPs, FNs, and FPs.

\subsection{Final results} 

We now present the results of ASID-L with the specific choices of S/N cut-off and LoG parameters introduced in the previous sections. At the same time, we compare the results with SExtractor, one of the most applied algorithms for source localization in optical images. The SExtractor results were obtained with the default parameters; the only exceptions were $DETECT\_MINAREA=3$ and $BACK\_SIZE=60$.

In Table \ref{table: Comparison}, for the Omega Cen test set, we compare the number of sources per square arcminute in the Gaia EDR3 catalog, and the sources localized by SExtractor and ASID-L.

\begin{table}[h]
\caption{Comparison of detected sources}             
\label{table: Comparison}      
\centering                          
\begin{tabular}{c c c c c c}        
\hline\hline                 
Method & TP & FN & FP & Dice  \\    
\hline                        
Gaia & 36.958 & - & - & - \\
SExtractor & 14.634 & 22.324 & 0.178 & 0.5653  \\
ASID-L & 27.001 & 9.957 & 2.918 & 0.8075 \\

\hline                                   
\end{tabular}
\tablefoot{Number of TPs, FNs, and FPs per square arcminute in the 165 patches of the Omega Cen test set for Gaia, Sextractor, and ASID-L.}
\end{table}

\begin{figure}[ht!]
   \centering
   \includegraphics[width=.45\textwidth]{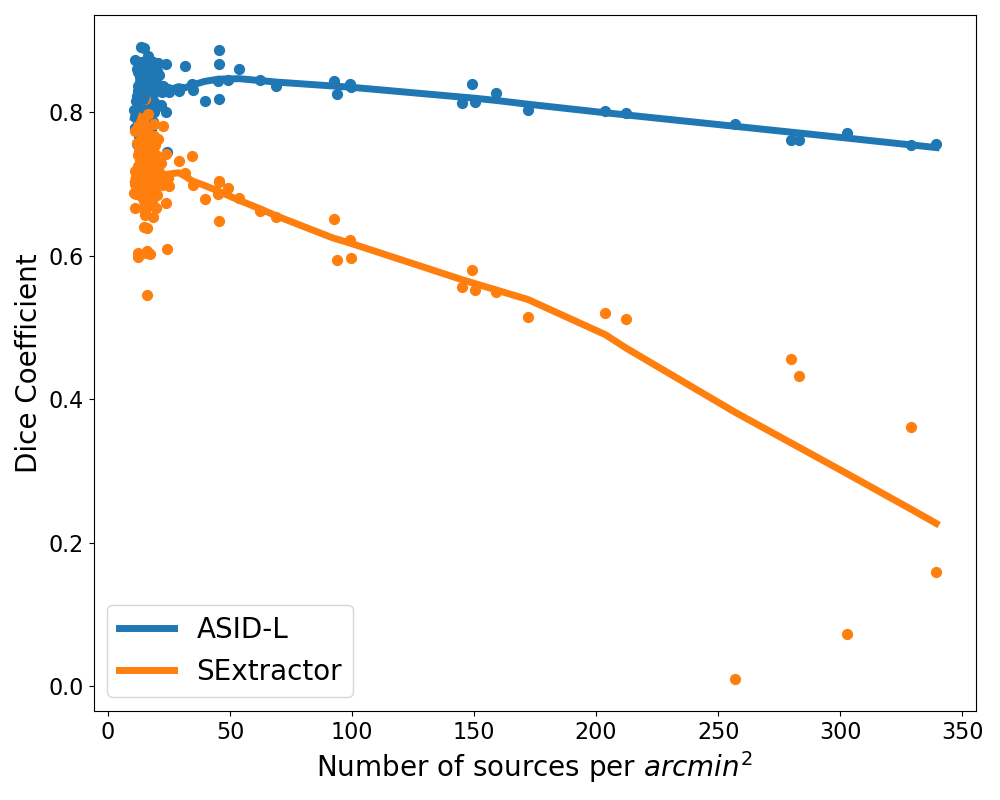}
   \caption{Dice coefficient of the Omega Cen test set patches as a function of the number of sources per square arcminute in each patch (ASID-L in blue and SExtractor in orange).}
    \label{fig: dice coefficient}
\end{figure}

\noindent
Gaia has the highest number of sources by far, and it is what we use as the ground truth for all the comparisons. However, the Gaia catalog is not perfect; it does not include small galaxies and might include sources that are not visible in the optical images for the reasons explained in Section \ref{Training set}. \\
For the Omega Cen test set, SExtractor ideally has a few FPs per square arcminute; however, the overall number of TPs is low, resulting in a Dice coefficient of $0.5653$. ASID-L, on the contrary, has an higher number of FPs, but it correctly localizes almost twice as many sources with respect to SExtractor, resulting in a Dice coefficient of $0.8075$. \\
We now evaluate the reason behind this substantial difference in the results of the two methods. In Fig. \ref{fig: dice coefficient}, we show the Dice coefficient, as in Eq. \ref{eqn:Dice coeff2}, for all 165 patches of the test set belonging to the Omega Cen globular cluster field.


\noindent
ASID-L scores are between $0.75$ and $0.9$ for any number of sources in the patches, proving that a high number of sources is not a concern for the method. ASID-L recovers more sources than SExtractor, although the influence of the FPs in low-density regions affects the resulting Dice coefficient value, as can be seen at the top left of the plot. SExtractor, instead, suffers in very crowded regions: the higher the number of sources, the lower its Dice coefficient. 

\begin{figure}[h!]
     \centering
     \begin{subfigure}[b]{0.24\textwidth}
         \centering
         \includegraphics[width=\textwidth]{Plots/OpticalImagePatch2.png}
         \caption{Uncrowded}
         
     \end{subfigure}
     \hfill
     \begin{subfigure}[b]{0.24\textwidth}
         \centering
         \includegraphics[width=\textwidth]{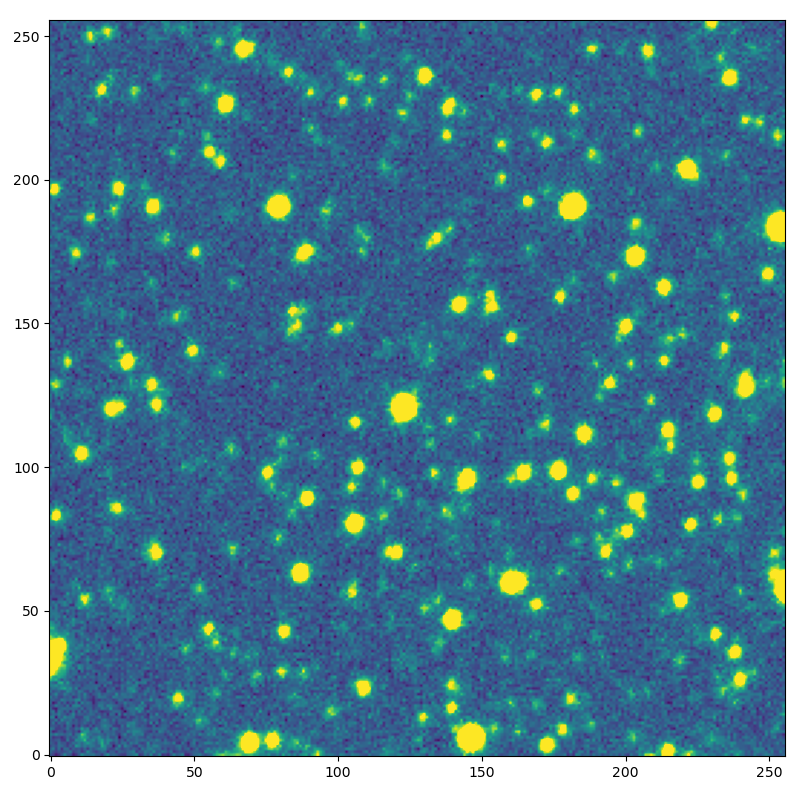}
         \caption{Crowded}
    \end{subfigure}
     
\caption{Optical patches with two different densities of sources.}
\label{fig: crowded/not crowded}
\end{figure}

\begin{figure}[h!]
     \centering
     \begin{subfigure}[b]{0.24\textwidth}
         \centering
         \includegraphics[width=\textwidth]{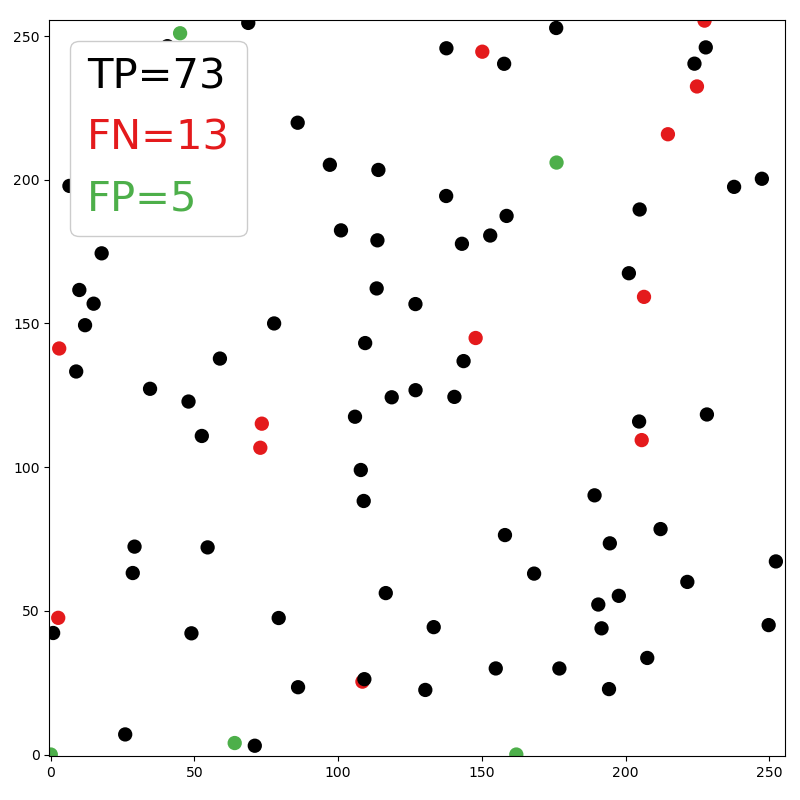}
         \caption{ASID-L}
         
     \end{subfigure}
     \hfill
     \begin{subfigure}[b]{0.24\textwidth}
         \centering
         \includegraphics[width=\textwidth]{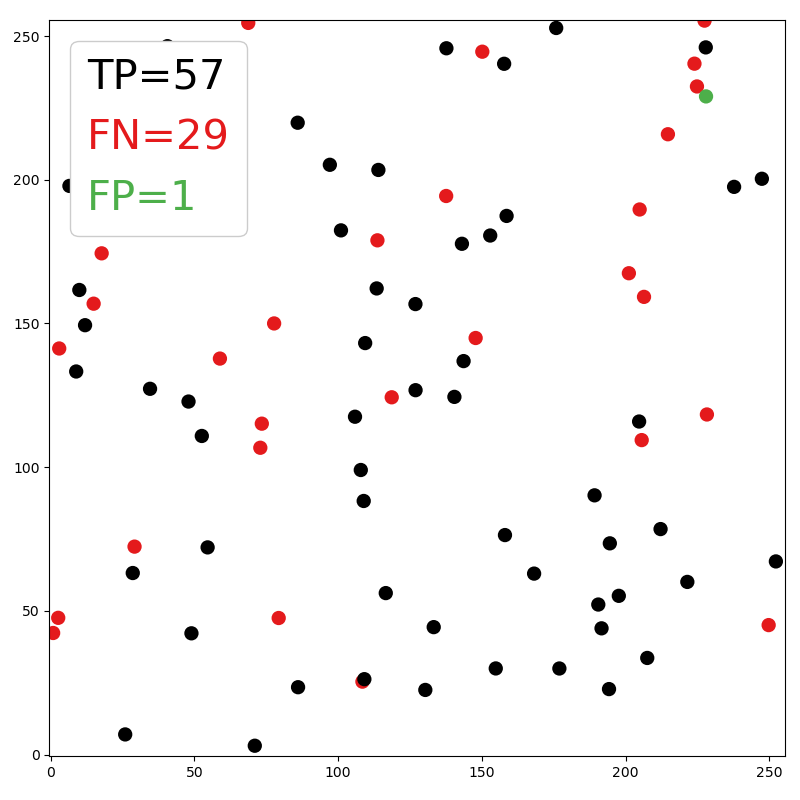}
         \caption{SExtractor}
     \end{subfigure}
\caption{Comparison between ASID-L and SExtractor with respect to GAIA EDR3 for the uncrowded patch in Fig. \ref{fig: crowded/not crowded}.}
\label{fig: uncrowded comparison}
\end{figure}

\begin{figure}[h!]
     \centering
     \begin{subfigure}[b]{0.24\textwidth}
         \centering
         \includegraphics[width=\textwidth]{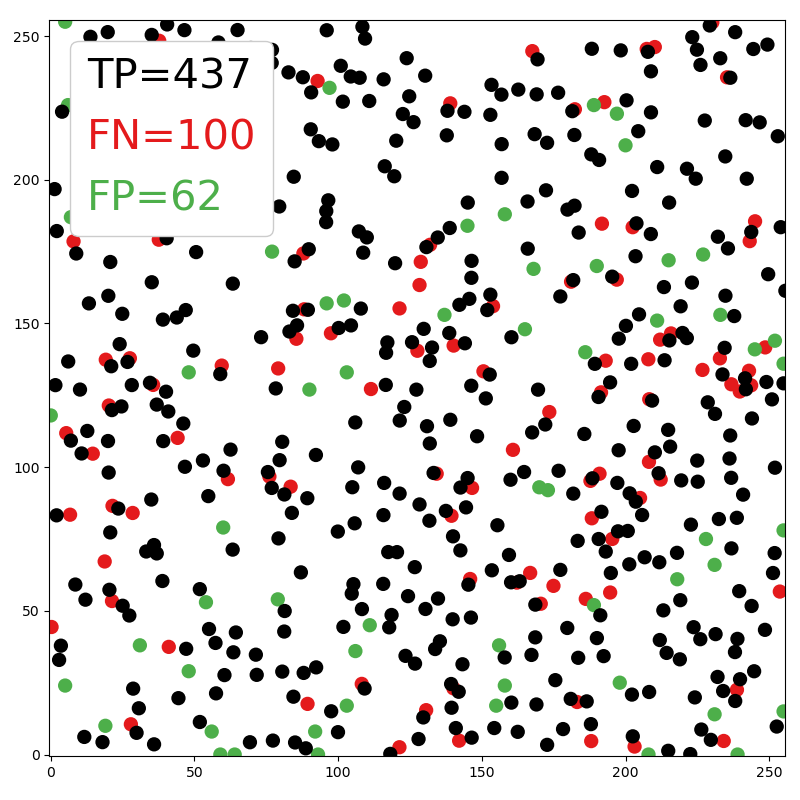}
         \caption{ASID-L}
         
     \end{subfigure}
     \hfill
     \begin{subfigure}[b]{0.24\textwidth}
         \centering
         \includegraphics[width=\textwidth]{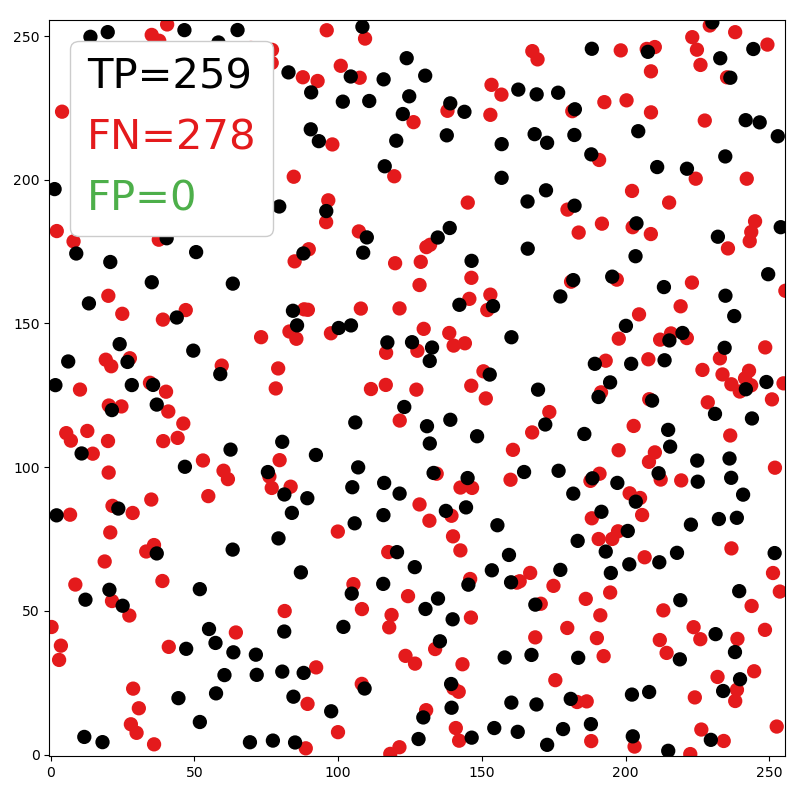}
         \caption{SExtractor}
     \end{subfigure}
\caption{Comparison between ASID-L and SExtractor with respect to GAIA EDR3 for the crowded patch in Fig. \ref{fig: crowded/not crowded}.}
\label{fig: crowded comparison}
\end{figure}

For the two patches of Fig. \ref{fig: crowded/not crowded}, Fig. \ref{fig: uncrowded comparison} and Fig. \ref{fig: crowded comparison} show the results of ASID-L and SExtractor with respect to Gaia EDR3 in terms of TP, FN, and FP.

\noindent
In uncrowded regions ASID-L recovers a few sources more than SExtractor, while for very crowded regions (as in Fig. \ref{fig: crowded comparison}), ASID-L localizes many more sources.

\section{Speed and additional features}

With the increased capabilities of many telescopes, large amounts of data will have to be processed at a staggeringly increasing rate. Thus, there is a need for computationally efficient methods that not only can keep up with this trend, but can also help to reduce the carbon footprint of this process. 

We now evaluate the processing time of ASID-L and SExtractor on an Alienware Area 51M, Intel Core i9-9900K, 32GB DDR4/2400, Nvidia GeForce RTX 2080.

We estimated the processing time of SExtractor in the Python library Source Extractor and Photometry (SEP, \citealp{Barbary2016, Barbary2017}). SEP uses the same core algorithms of SExtractor, it is written in C, and it has a Python module to wrap it in a Pythonic API. This additional step makes SEP slower than SExtractor, but within acceptable limits.

ASID-L was also tested in Python. It was developed with Tensorflow and Scikit-Image, runs on GPU, and was parallelized on multiple cores, although the use of a GPU or parallelization is not needed. 

In Table \ref{table:5}, we evaluate the time performance of both methods on 3 2560 $\times$ 2560 pixels (580 square arcminutes) MeerLICHT images with different densities of sources.

\begin{table}[h]
\caption{Computation time comparison.}             
\label{table:5}      
\centering                          
\begin{tabular}{c c c}        
\hline\hline                 
$\#$ sources & SExtractor [s] & ASID-L [s]   \\    
\hline                        
  10    & 1.9687 & 0.2969 \\
  75    & 10.2188 & 0.2812 \\
  250   & 25.2510 & 0.2969 \\
\hline                                   
\end{tabular}
\tablefoot{
Time in seconds to process each image with different source densities per square arcminute.
}
\end{table}


\noindent
The striking difference between ASID-L and SExtractor is that the ASID-L processing time does not depend on the number of sources in the images; SExtractor, on the contrary, does.
For fields with approximately ten sources per square arcminute ASID-L runs seven times faster than SEP; however, SEP provides a set of additional information for each source that makes it preferable for the accurate exploration of uncrowded fields. 
SExtractor is a great tool, but it also heavily depends on the input parameters, which makes it unsuitable for the automatic localization of sources. ASID-L, on the contrary, does not need any prior information about the field and can be applied without supervision in a live-stream manner. ASID-L not only speeds up the source localization task, but its additional features make the preprocessing of the optical images unnecessary, which increases the time gain and reduces the carbon footprint of the entire process.

ASID-L is trained and can thus predict on images where the background has not been subtracted and where differences between the channels of the CCD have not been corrected, effectively removing these steps from the preprocessing completely.
ASID-L can also discern between real sources and a series of artifacts, such as diffraction spikes and cosmic rays. 
The cosmic-ray removal was previously performed using the Astroscrappy (https://github.com/astropy/astroscrappy) implementation of L.A. Cosmic \citep{vanDokkum2001,vanDokkum2012} and required a few minutes per MeerLICHT/BlackGEM \citep{Groot2019b} image. This is now taken care of directly during the prediction step.

\begin{figure}[h!]
     \centering
     \begin{subfigure}[b]{0.24\textwidth}
         \centering
         \includegraphics[width=\textwidth]{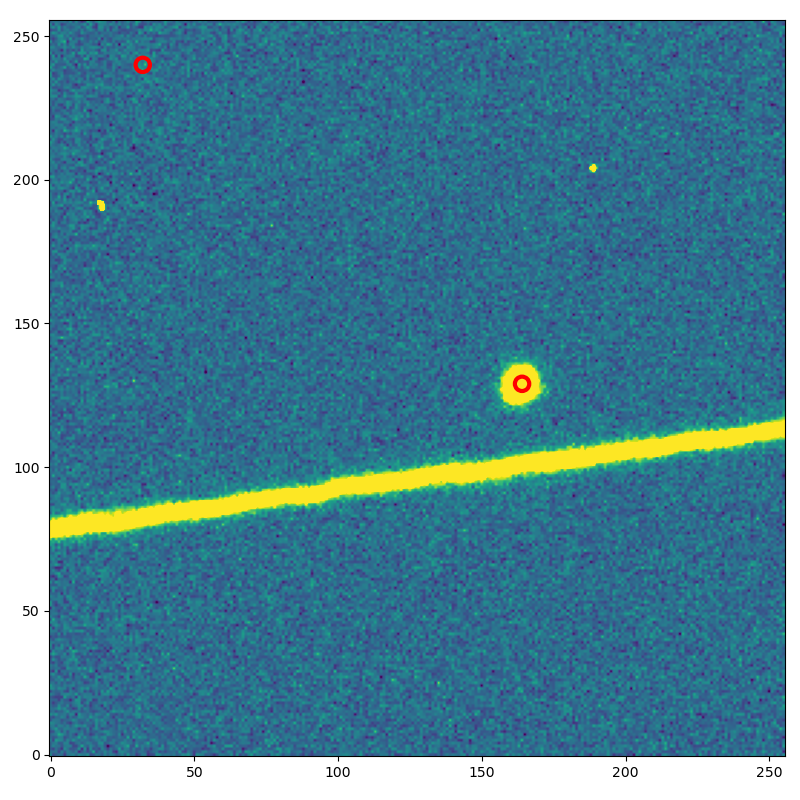}
         \caption{Cosmic rays and satellite}
         
     \end{subfigure}
     \hfill
     \begin{subfigure}[b]{0.24\textwidth}
         \centering
         \includegraphics[width=\textwidth]{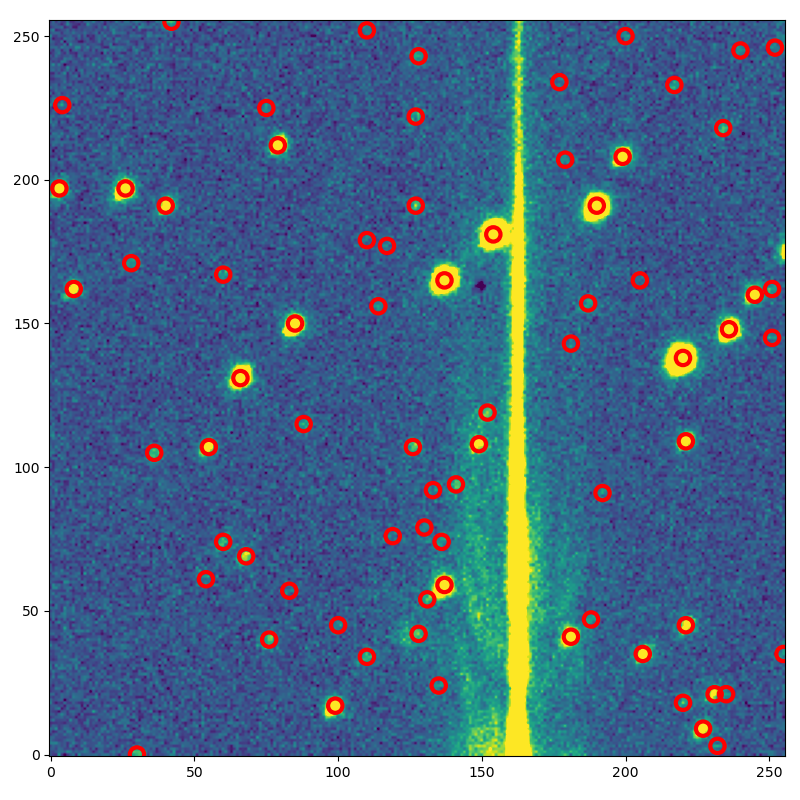}
         \caption{Diffraction spikes}
     \end{subfigure}
\caption{Sources localized by ASID-L (red circle) superimposed on two optical images in the presence of multiple artifacts.}
\label{fig: satellite and cosmic}
\end{figure}

Last but not least, ASID-L can recognize satellite trails and correctly discard them. With the increasing number of low-Earth-orbit (LEO) satellites such as SpaceX's Starlink, this is a vital feature. It has been calculated that once the size of the Starlink constellation reaches 10,000, essentially all ZTF images taken during twilight may be affected \citep{Mroz2022}. 
In Fig. \ref{fig: satellite and cosmic}, we show the prediction of ASID-L on two images affected by cosmic rays, satellite trails and diffraction spikes.


\section{Transfer learning and application to different telescopes}
\label{section: transfer}

\begin{figure}[ht!]
   \centering
   \includegraphics[width=.5\textwidth]{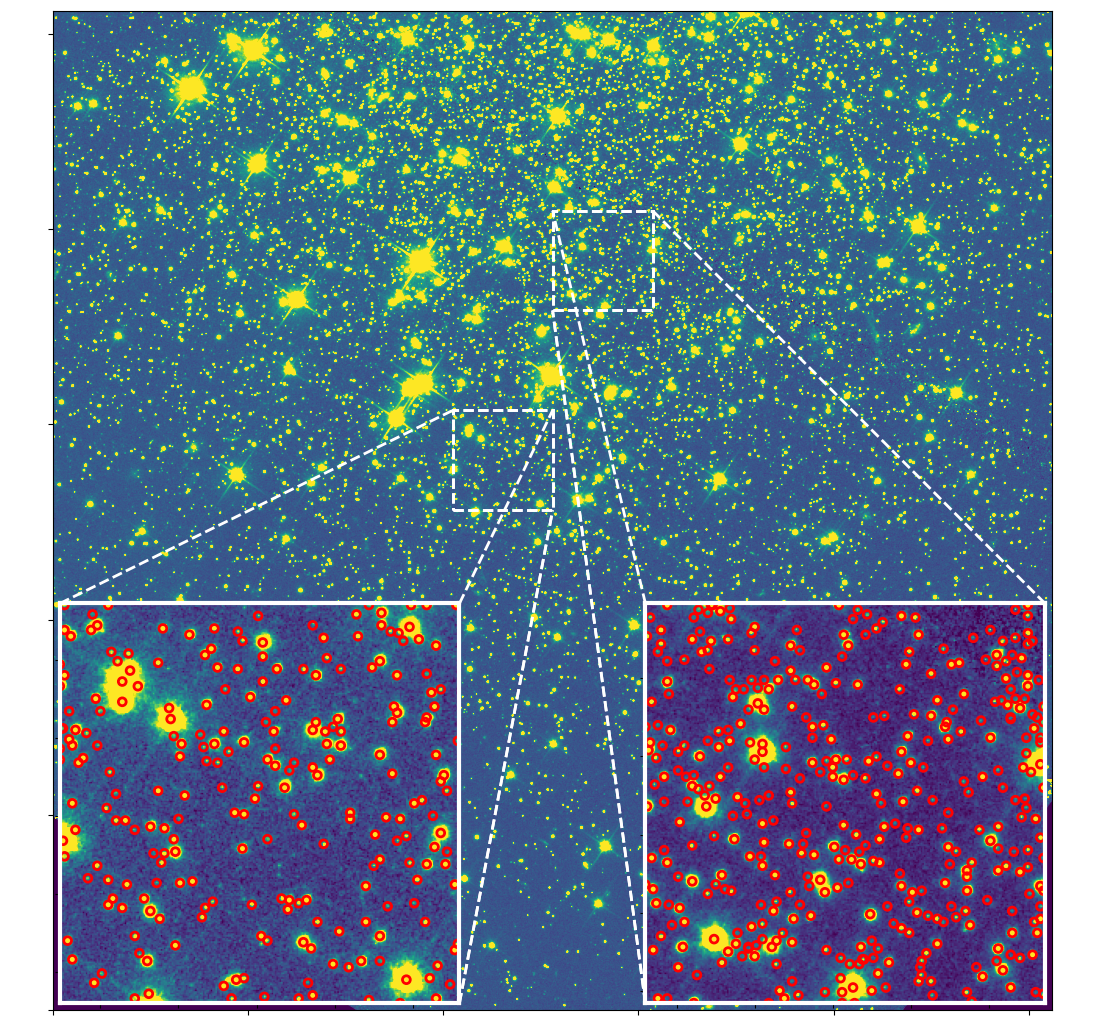}
   \caption{Star cluster image retrieved from the Hubble Space Telescope archive (GO-10396, PI: J.S. Gallagher). In red are the sources localized by ASID-L.}
     \label{HSTField10396}
\end{figure}

\begin{figure}[ht!]
   \centering
   \includegraphics[width=.5\textwidth]{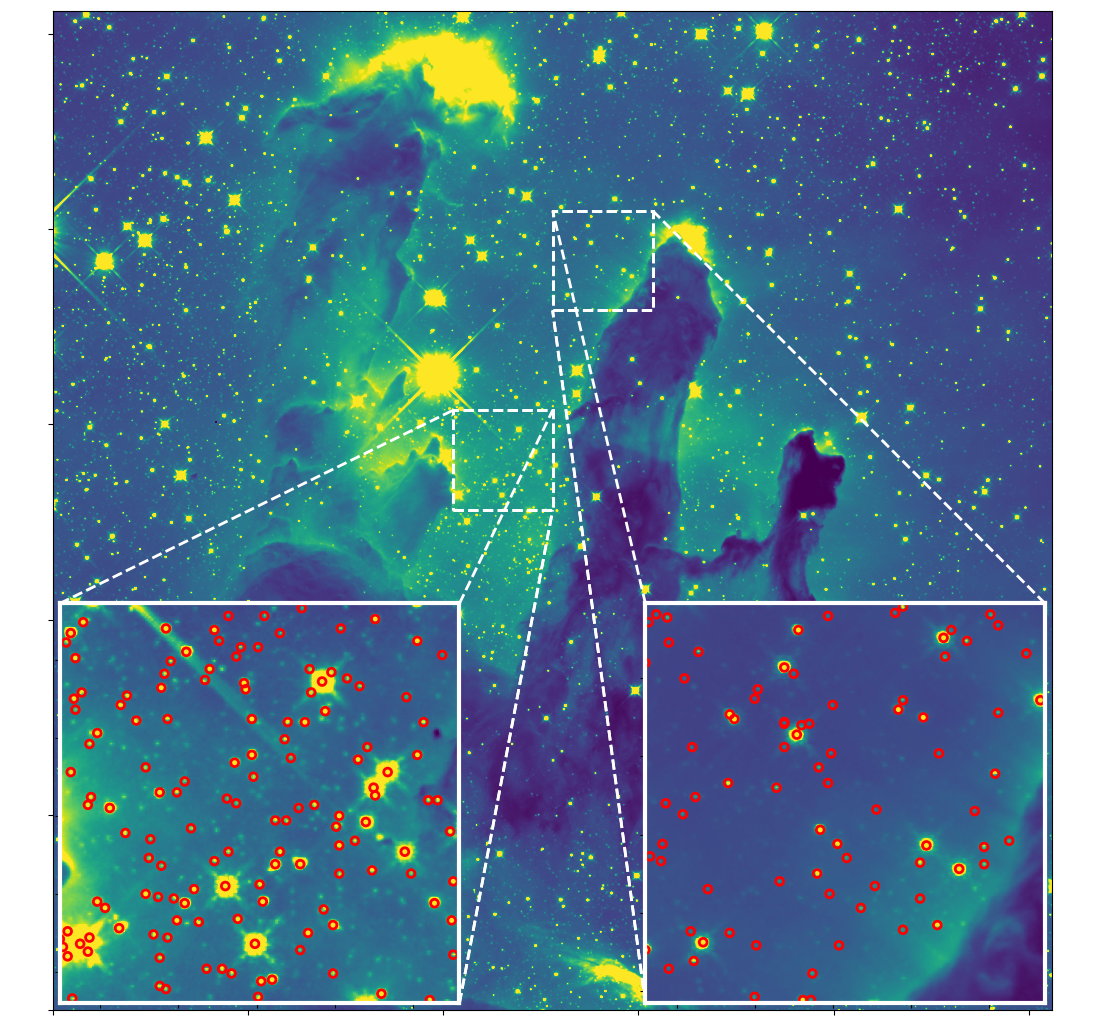}
   \caption{Messier 16 (Eagle Nebula) image retrieved from the Hubble Space Telescope archive. In red are the sources localized by ASID-L.}
     \label{HSTfieldM16}
\end{figure}

An open question that we want to address in the future is how the resolution of the images affects the localization results. A promising first test can be found in Figs. \ref{HSTField10396} and \ref{HSTfieldM16}, where we applied ASID-L, trained on MeerLICHT images, to images from the Hubble Space Telescope. The HST has a full width at half maximum (FWHM) PSF of about $0.11$ arcseconds, much better than the $2$--$3$ arcseconds of MeerLICHT.

\begin{figure}[ht!]
   \centering
   \includegraphics[width=.5\textwidth]{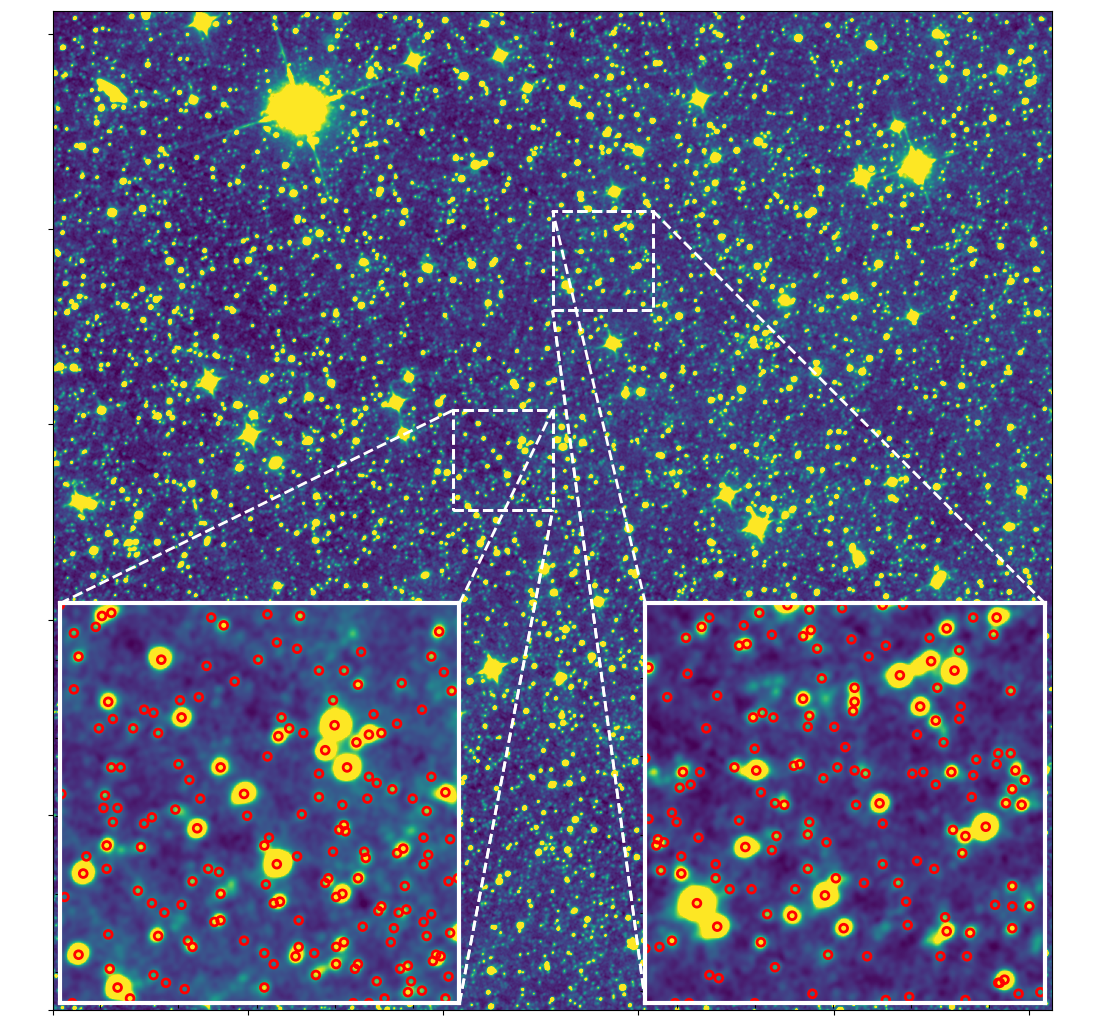}
   \caption{Infrared image of the NGC31 field taken with the WISE space telescope. In red are the sources localized by ASID-L.}
     \label{WISEField}
\end{figure}

Although this is an early study, it appears that ASID-L is capable of localizing many sources without the need to retrain the U-Net on HST images. The main difference between MeerLICHT and HST, the resolution of the images, does not seem to affect the results of the method. However, there may be artifacts in HST images that differ from those ASID-L ever encountered in MeerLICHT, one example being the diffraction spikes that are much brighter for HST. \\
Moreover, in Fig. \ref{HSTfieldM16} we can see that some small sources are not localized by ASID-L; two possible reasons could be the peculiar background or the fact that in MeerLICHT sources of that size are most likely cosmic rays that ASID-L is trained to automatically remove.\\
ASID-L was created for optical images, but we are also interested in its performance for different parts of the EM spectrum. In Fig. \ref{WISEField} we show the result of ASID-L applied to an infrared image from the Wide-field Infrared Survey Explorer (WISE, \citealp{Wright2010}) space telescope.
ASID-L seems to hold localization power also in the infrared; although the background is quite different from that of the MeerLICHT images, many sources are localized. A deeper exploration of these images with a ground truth catalog, like we did with Gaia EDR3, will give us many insights into what the differences are in applying ASID-L to different parts of the EM spectrum.

A great result for the future would be proving that ASID-L can be applied to different telescopes without the need of retraining for each specific instrument, effectively opening a path toward transfer learning for a broad range of telescopes.

\section{Conclusions}
In this paper we presented the building blocks of ASID-L and applied it to real images taken with the MeerLICHT telescope. 
Born from the combination of a U-Net network and a computer vision tool called Laplacian of Gaussian, ASID-L has been trained and evaluated with the Gaia EDR3 catalog. The result is a clean framework for significantly increasing the speed and accuracy of optical source localization in any field, crowded or not.

Because it does not depend on any prior knowledge, ASID-L can be used on archival images or live as soon as a telescope takes the image. An iterative use of ASID-L can be applied to the search for specific objects like transients and variable stars. ASID-L not only identifies faint sources, it also handles a few well-known optical image problems: cosmic rays, diffraction spikes, and artificial trails. 

The dataset used in this paper for training, test, and validation set is available on Zenodo (\citealp{Stoppa2022_a} \footnote{\url{https://doi.org/10.5281/zenodo.5902893}}). ASID-L is directly accessible on GitHub and Zenodo (\citealp{Stoppa2022_b, Stoppa_c} \footnote{\url{https://doi.org/10.5281/zenodo.5938341}}), and will be included in the pipeline of the MeerLICHT telescope.

Our next project will be to expand ASID-L and make its output broader by including additional information about the localized sources.
The current output of ASID-L is suitable to be used as input for additional networks; one example is cutting $32 \times 32$ pixel patches around each localized source and applying a classifier to discern between different types of sources or simply to discern between true positives and false positives and improving the performance of ASID-L.
Another deep learning algorithm can be applied to each localized source to extract their features such as flux, PSF, ellipticity, the pixels they occupy, and a more accurate estimate of the center. All these additional steps and the compatibility with multiple telescopes will further enhance the competitiveness of ASID-L in the astronomical community.

\begin{acknowledgements}
      F. S. and G. N. acknowledge support from the Dutch Science Foundation NWO.\\
      S. B. and G. Z. acknowledge the financial support from the Slovenian Research Agency (grants P1-0031, I0-0033 and J1-1700).\\
      R. Rda acknowledges support from the Ministerio de Ciencia e Innovación (PID2020-113644GB-I00).\\
      P.J. G. is supported by NRF SARChI (grant 111692).\\
      The MeerLICHT telescope is a collaboration between Radboud University, the University of Cape Town, the South African Astronomical Observatory, the University of Oxford, the University of Manchester and the University of Amsterdam, and supported by the NWO and NRF Funding agencies.\\
      We thank the anonymous reviewer whose comments and suggestions helped to improve and clarify this paper.
\end{acknowledgements}

\bibliographystyle{aa}
\bibliography{bibliography}

\end{document}